\documentclass[12pt]{article}
\usepackage{graphicx}
\usepackage{amsmath}
\usepackage{amssymb}
\usepackage{cite}
\def\Title#1{\begin{center} {\Large\bf #1 } \end{center}}
\def\Author#1{\begin{center}{ \sc #1} \end{center}}
\def\Address#1{\begin{center}{ \it #1} \end{center}}

\newcommand\pubblock{\rightline{\begin{tabular}{l} \pubnumber\\
        \pubdate\\ \hepnumber \end{tabular}}}
\newcommand\pubnumber{UAB-FT-546\\PSI-PR-03-11\\UB-ECM-PF-03/13}
\newcommand\pubdate{July 2003}
\newcommand\hepnumber{hep-ph/0307144}

\newcommand{\tb}{\tan\beta}

\newcommand{\stbt}{\sin2\beta}
\newcommand{\ctbt}{\cos2\beta}

\newcommand{\ta}{\tan\alpha}

\newcommand{\sbma}{\sin (\beta-\alpha)}
\newcommand{\cbma}{\cos (\beta-\alpha)}

\newcommand{\mw}{M_W}

\newcommand{\mh}{m_{h^0}}
\newcommand{\mH}{m_{H^0}}
\newcommand{\mHp}{m_{H^\pm}}
\newcommand{\mA}{m_{A^0}}

\newcommand{\mHs}{m^2_{H^0}}
\newcommand{\mHps}{m^2_{H^\pm}}
\newcommand{\mAs}{m^2_{A^0}}

\begin{document}

\pubblock

\vfill
\def\thefootnote{\fnsymbol{footnote}}
\Title{Higgs Boson Flavor-Changing Neutral Decays into Top Quark
  in a General Two-Higgs-Doublet Model} \Author{Santi
  B{\'e}jar$^{a,b}$, Jaume Guasch$^{c}$, Joan Sol{\`a}$^{d,b}$}
\Address{
  $^{a}$\textsl{Grup de F{\'\i}sica Te{\`o}rica,
    Universitat Aut{\`o}noma de Barcelona,\\
    E-08193, Bellaterra,
    Barcelona, Catalonia, Spain\\ }
  $^{b}$\textsl{
    Institut de F{\'\i}sica d'Altes Energies, Universitat Aut{\`o}noma de Barcelona,\\
    E-08193, Bellaterra, Barcelona, Catalonia, Spain\\ }
  $^{c}$\textsl{Theory Group LTP, Paul Scherrer Institut, CH-5232 Villigen
    PSI, Switzerland\\ }
  $^{d}$\textsl{Departament d'Estructura i Constituents de la
    Mat{\`e}ria, Universitat de Barcelona,\\
    E-08028, Diagonal 647, Barcelona, Catalonia, Spain}
} \vspace{1cm}
\begin{abstract}
    Higgs boson decays mediated by flavor changing neutral currents
    (FCNC) are very much suppressed in the Standard Model, at the
    level of $10^{-15}$ for Higgs boson masses of a few hundred $GeV$.
    Therefore, any experimental vestige of them would immediately
    call for new physics. In this paper we consider the FCNC decays
    of Higgs bosons into a top quark in a general two-Higgs-doublet
    model {(2HDM)}. The isolated top quark signature, unbalanced by any other
    heavy particle, should help to identify the potential FCNC events
    much more than any other final state. We compute the maximum
    branching ratios and the number of FCNC Higgs boson decay events
    {at} the LHC collider at CERN. The most favorable mode for production
    and subsequent FCNC decay is the lightest CP-even state in the
    Type II {2HDM}, followed by the other CP-even
    state, if it is not very heavy, whereas the CP-odd mode can never
    be sufficiently enhanced. Our calculation shows that the
    branching ratios of the CP-even states may reach $10^{-5}$, and that several
    hundred events could be collected in the highest luminosity runs
    of the LHC. We also point out some strategies to use these  FCNC decays
    as a handle to discriminate {between} 2HDM and
    {supersymmetric} Higgs bosons.

\end{abstract}


\baselineskip=5.6mm

\section{Introduction}

At the tree-level there are no {Flavor Changing Neutral {Currents}
(FCNC)} processes in the {Standard Model (SM)}, and at one-loop
they are induced by charged-current interactions, which are
GIM-suppressed~\cite{GIM}. Letting aside the meson-meson
oscillations, such as $K^0-\bar{K}^0$ and $B^0-\bar{B}^0$, the
decay processes mediated by FCNC are also of high interest and are
strongly suppressed too. For instance, we have the radiative
B-meson decays, with a typical branching ratio $BR(b\rightarrow
s\,\gamma)\sim10^{-4}$. But we also have the FCNC decays with the
participation of the top quark as a physical field, which are by
far the most suppressed decay modes~\cite{LorenzoD,Eilam:1991zc}.
Indeed, the top quark decays into gauge bosons ($t\rightarrow
c\,V$;$\;V\equiv\gamma,Z,g$) are well known to be extremely rare
events in the SM. The branching ratios are, according to
Ref.\,\cite{Eilam:1991zc}: $\sim 5\times 10^{-13}$ for the photon,
slightly above $1\times10^{-13}$ for the $Z$-boson, and
$\sim4\times10^{-11}$ for the gluon channel, or even smaller
according to other estimates \,\cite{JASaavedra03}. Similarly, the
top quark decay into the SM Higgs boson, $H^{SM}$, is a very
unusual decay, typically $BR(t\rightarrow c\,H^{SM})\sim
10^{-14}$\,\cite{Mele}. However, when considering physics beyond
the SM, new horizons of possibilities open up which may radically
change the pessimistic prospects for FCNC decays involving a Higgs
boson and the top quark. For example, in Ref.\cite{Divitiis} it
was shown that the vector boson modes can be highly enhanced
within the context of the Minimal Supersymmetric Standard Model
(MSSM) \,\cite{MSSMreps}. This fact was also dealt with in great
detail in Ref. \cite{GuaschNP1} where in addition a dedicated
study was presented of the FCNC top quark decays into the various
Higgs bosons of the MSSM (see also \cite{Yuan}), showing that
these can be the most favored FCNC top quark decays -- above the
expectations on the gluon mode $t\rightarrow c\,g$. A similar
study was performed for the FCNC top quark decays into Higgs
bosons in a general two-Higgs-doublet model (2HDM)\cite{BejarNP1}.

The low observed rates of FCNC processes among the SM fields
strongly suggest that their FCNC couplings must be suppressed at
the tree-level in any extension of the SM, in particular in the
2HDM models. As is well known, there are two canonical strategies
to get rid of these tree-level couplings in that kind of
models~\cite{Hunter}. In Type~I 2HDM (also denoted 2HDM~I) one
Higgs doublet, $\Phi_{1}$, does not couple to fermions at all and
the other Higgs doublet, $\Phi_{2}$, couples to fermions in the
same manner as in the SM. In contrast, in Type~II 2HDM (also
denoted 2HDM~II) one Higgs doublet, $\Phi_{1}$, couples to down
quarks (but not to up quarks) while $\Phi_{2}$ does the other way
around. Such a coupling pattern is automatically realized in the
framework of supersymmetry (SUSY), in particular in the MSSM, but
it can also be arranged in non-supersymmetric extensions if we
impose a discrete symmetry, e.g. $\Phi _{1}\rightarrow-\Phi_{1}$
and $\Phi_{2}\rightarrow+\Phi_{2}$ (or vice versa) plus a suitable
transformation for the right-handed quark fields. In an analysis
of FCNC top quark decays in 2HDM extensions of the SM
\,\cite{BejarNP1,BejarRADCOR} it was proven that while the
maximum rates for $t\rightarrow c\,g$ were one order of magnitude
more favorable in the MSSM than in the 2HDM, the corresponding
rates for $t\rightarrow c\,h^0$ were comparable both for the MSSM
and the general 2HDM, namely up to the $10^{-4}$ level and should
therefore be visible both at the LHC and the
LC\,\cite{BejarNP1,BejarRADCOR,Saavedra}.

Similarly, one may wonder whether the FCNC decays of the Higgs
bosons themselves can be of some relevance. Obviously the
situation with the SM Higgs is essentially hopeless, so again we
have to move to physics beyond the SM. Some work on these decays,
performed in various contexts including the MSSM, shows that
these effects can be important \,\cite{Hou,Curiel,Demir,
Brignole, unpublished}. This could be expected, at least for
heavy quarks in the MSSM, from the general SUSY study (including
both strong and electroweak supersymmetric effects) of the FCNC
vertices $h\,t\,c$ ($h=h^0\,,H^0\,,A^0$) made in  Ref.
\cite{GuaschNP1}. However, other frameworks could perhaps be
equally advantageous. Here we are particularly interested in the
FCNC Higgs decay modes into top quark within a general 2HDM,
which have not been studied anywhere in the literature to our
knowledge. It means we restrict to Higgs bosons heavier than
$m_t$. From the above considerations, and most particularly on
the basis of the detailed results obtained in
\cite{BejarNP1,BejarRADCOR} one may expect that some of the
decays of the Higgs bosons
\begin{equation}\label{hFCNC}
    h\rightarrow t\,\bar{c}\,,\ \  \ \ h\rightarrow \bar{t}\,c\,\ \
    \ \ \ (h=h^0\,,H^0\,,A^0)
\end{equation}
in a general 2HDM can be substantially enhanced and perhaps can be
pushed up to the visible level, particularly for $h^0$ which is
the lightest CP-even spinless state in these
models\,\cite{Hunter}. This possibility can be of great relevance
on several grounds. {On the one} hand the severe degree of
suppression of the FCNC Higgs decay in the SM obviously implies
that any experimental sign of Higgs-like FCNC decay (\ref{hFCNC})
would be instant evidence of physics beyond the SM. {On the other
hand}, the presence of an isolated top quark in the final state,
unbalanced by any other heavy particle, is an unmistakable
carrier of the FCNC\,  signature. {Finally}, the study of the
maximum FCNC rates for the top quark modes (\ref{hFCNC}) within
the 2HDM, which is the simplest non-trivial extension of the SM,
should serve as a fiducial result from which more complicated
extensions of the SM can be referred to. Therefore, we believe
there are founded reasons to perform a thorough study of the FCNC
Higgs decays in minimal extensions of the Higgs sector of the SM
and see whether they can be of any help to discover new physics.

The paper is organized as follows. In Section 2 {we summarize the
2HDM interactions most relevant for our study and} estimate the
expected FCNC rates of the Higgs decays in the SM and the general
2HDM. In Section 3 a detailed numerical analysis of the one-loop
calculations of the FCNC decay widths and production rates of
FCNC Higgs events is presented. Finally, in Section 4 we discuss
the reach of our results and its phenomenological implications,
and deliver our conclusions.

\section{Expected branching ratios in the SM and the 2HDM}

Before presenting the detailed numerical results of our
calculation, it may be instructive to {estimate} the typical
expected widths and branching ratios ($BR$) both for the SM decay
$H^{SM}\rightarrow t\,\bar{c}\ $ and the non-standard decays
(\ref{hFCNC}) in a general 2HDM. This should be especially useful
in this kind of rare processes, which in the strict context of
the SM are many orders of magnitude out of the accessible range.
Therefore, one expects to be able to grossly reproduce the order
of magnitude from simple physical considerations based on
dimensional analysis, power counting, CKM matrix elements and
dynamical features. By the same token it should be possible to
{guess at the} potential enhancement factors in the 2HDM extension
of the SM. In fact, guided by the previous criteria the FCNC decay
width of the SM Higgs of mass $m_H$ into top quark is expected to
be of order
\begin{equation}
    \Gamma(H^{SM}\rightarrow
    t\,\bar{c})\sim\left(\frac{1}{16\pi^2}\right)^2\,
    |V_{tb}^{\ast}V_{bc}|^{2}\, \alpha_{W}^3\, \,m_{H}\left(
        \lambda_b^{SM}\right) ^{4}
    \sim\left(\frac{|V_{bc}|}{16\pi^2}\right)^2\,\alpha_{W}\ G_F^2
    \,m_{H}\,m_b^4\,,\label{estimateSM1}
\end{equation}
where $G_F$ is Fermi's constant and $\alpha_{W}=g^2/4\pi$, $g$
being the $SU(2)_L$ weak gauge coupling. We have approximated the
loop form factor by just a constant prefactor. Notice the
presence of $\lambda_b^{SM}\sim m_b/M_W$, which is the SM Yukawa
coupling of the bottom quark in units of $g$. The fourth power of
$\lambda_b^{SM}$ in (\ref{estimateSM1}) gives the non-trivial
suppression factor reminiscent of  the GIM mechanism after
summing over flavors. Since we are maximizing our estimation, a
missing function related to kinematics and polarization sums,
$F(m_{t}/m_{H})\sim(1-m_{t}^{2}/m_{H}^{2})^{2}$, has been
approximated to one. To obtain the (maximized!) branching ratio it
suffices to divide the previous result by
$\Gamma(H^{SM}\rightarrow b\,\bar{b})\sim\alpha_W\,\left(
    \lambda_b^{SM}\right) ^{2}\,m_H\sim G_F\,m_H\,m_b^2\ $  to obtain
\begin{equation}
    BR(H^{SM}\rightarrow t\,\bar{c})
    \sim\left(\frac{|V_{bc}|}{16\pi^2}\right)^2\,\alpha_{W}\
    G_F\,m_b^2\sim 10^{-13}\,,\label{estimateBRSM}
\end{equation}
with $V_{bc}=0.04$, $m_b=5\,GeV$. In general this $BR$ will be
even smaller, specially for higher Higgs boson masses ($m_H>2\,M_W
$) for which the vector boson Higgs decay modes $H^{SM}\rightarrow
W^+\,W^-(Z\,Z)$ can be kinematically available and become
dominant. In this case it is easy to see that
$BR(H^{SM}\rightarrow t\,\bar{c})$ will be suppressed by an
additional factor of $m_b^2/m_H^2$, which amounts at the very
least to two additional orders of magnitude suppression, bringing
it to a level of less than $10^{-15}$. Already the optimized
branching ratio (\ref{estimateBRSM}) will remain invisible to all
foreseeable accelerators in the future! {To obtain the
corresponding maximized estimation for the 2HDM we remind the
following basic interaction Lagrangians within the general 2HDM.
The charged Higgs interactions with fermions read}
\begin{equation}
\mathcal{L}_{Htb}^{(j)}=\frac{gV_{tb}}{\sqrt{2}\,M_{W}}\,H^{-}\overline{b}\,%
\left[ m_{t}\cot \beta \,P_{R}+m_{b}\,a_{j}\,P_{L}\right]
\,t+h.c. \label{Htb}
\end{equation}
{We use third-quark-family notation as a generic one;
$P_{L,R}=(1/2)(1\mp \gamma _{5})$
are the chiral projection operators on left- and right-handed fermions; $%
j=I,II$ runs over Type~I and Type~II 2HDM's. A most relevant
parameter here is $\tb$, the ratio between the vacuum expectation
values of the two Higgs doublet models\,\cite{Hunter}. In the
Lagrangian above,  the parameter $a_{j}$ is such that
$a_{I}=-\cot \beta $ and $a_{II}=+\tan \beta $}. For the neutral
Higgs interactions, \ the necessary pieces of the Lagrangian are
the following:
\begin{eqnarray}
\mathcal{L}_{hqq}^{(j)} &=&\frac{-g\,m_{b}}{2\,M_{W}\,\left\{
\begin{array}{c}
\sin \beta \\
\cos \beta
\end{array}
\right\} }\,\overline{b}\left[ h^{0}\,\left\{
\begin{array}{c}
\cos \alpha \\
-\sin \alpha
\end{array}
\right\} +H^{0}\,\left\{
\begin{array}{c}
\sin \alpha \\
\cos \alpha
\end{array}
\right\} \right] \,b+\frac{i\,g\,m_{b}\,a_{j}}{2\,M_{W}}\,\overline{b}%
\,\gamma _{5}\,b\,A^{0}  \notag \\
&+&\frac{-g\,m_{t}}{2\,M_{W}\,\sin \beta }\,\overline{t}\left[
h^{0}\,\cos
\alpha +H^{0}\,\sin \alpha \right] \,t+\frac{i\,g\,m_{t}}{2\,M_{W}\tb}\,%
\overline{t}\,\gamma _{5}\,t\,A^{0}\,\,,  \label{Hff}
\end{eqnarray}
{where the upper row is for Type I models and the down row is for
Type II. Here $\alpha$ is the mixing angle in the CP-even Higgs
sector. Apart from the interactions with fermions, there is a set
of potentially very relevant trilinear self-interactions of Higgs
bosons in a general 2HDM. These are summarized in
Table~\ref{tab:trilineals} \footnote{Table~\ref{tab:trilineals}
extends the one in Ref.\,\cite{BejarNP1}, and therefore we also
assume here $\lambda_5=\lambda_6$ in the 2HDM Higgs
potential\,\cite{Hunter}. We have omitted the couplings with the
Goldstone bosons. Although the latter have been included in the
calculation, their potential enhancement is far less significant
than in the case of the Higgs trilinear couplings. }. Notice that
the trilinear couplings in Table~\ref{tab:trilineals} are common
to the 2HDM I and 2HDM II, and they are quite different from the
ones obtained in the MSSM, whose pure gauge structure is enforced
by the underlying supersymmetry.}
\begin{table}[tb]
    \centering
    \begin{tabular}{|c|c|}
        \hline
        $H^{\pm}H^{\mp}H^0$&$-\frac{g}{\mw\stbt}
        \left[(\mHps-\mAs+\frac{1}{2}\mHs)\stbt\cbma+\right.$\\
        &$\phantom{\frac{g}{\mw\stbt}}\left.+(\mAs-\mHs)\ctbt\sbma\right]$\\\hline
        $H^{\pm}H^{\mp}h^0$&$-\frac{g}{\mw\stbt}
        \left[(\mHp^2-\mA^2+\frac12\mh^2)\sin{2\beta}\sin(\beta-\alpha)+\right.$\\
        &$\phantom{\frac{g}{\mw\stbt}}
        \left.+(\mh^2-\mA^2)\,\cos{2\beta}\,\cos(\beta-\alpha)\right]$\\\hline
        $h^0h^0H^0$&$-\frac{g\,\cos(\beta-\alpha)}{2\,M_W\,\sin{2\beta}}\,
        \left[(2\,\mh^2+\mH^2)\,\sin{2\alpha}-\right.$\\
        &$\phantom{-\frac{g}{\mw\stbt}}
        \left.-\mA^2\,(3\sin{2\alpha}-\sin{2\beta})\right]$\\\hline
        $A^0A^0H^0$&$-\frac{g}{2\,M_{W}\sin{2\beta}}\,
        \left[\mH^2\,\sin{2\beta}\cos(\beta-\alpha)+\right.$\\
        &$\phantom{-\frac{g}{\mw\stbt}}
        \left.+2(\mH^2-\mA^2)\,\cos{2\beta}\,\sin(\beta-\alpha)\right]$\\\hline
        $A^0A^0h^0$&$-\frac{g}{2\,\mw\stbt}
        \left[\mh^2\,\sin{2\beta}\sin(\beta-\alpha)+\right.$\\
        &$\phantom{-\frac{g}{\mw\stbt}}
        \left.+2(\mh^2-\mA^2)\,\cos{2\beta}\,\cos(\beta-\alpha)\right]$\\\hline
    \end{tabular}
\caption{The needed trilinear self-couplings of the Higgs bosons
in the 2HDM within the framework of
Ref.\,\protect{\cite{BejarNP1}}. \label{tab:trilineals}}
\end{table}

Let us now first assume large $\tan\beta$ and {restrict to Type
II models. {From the interaction Lagrangians above it is clear
that we may replace} $\lambda_b^{SM}\rightarrow
\lambda_b^{SM}\,\tan\beta$ in the previous formulae {for the
partial width.} Moreover, the leading diagrams in the 2HDM
contain the trilinear Higgs couplings $\lambda_{H^+\,H^-\,h}$.
Therefore, the maximum $BR$ associated to the FCNC decays
(\ref{hFCNC}) in a general 2HDM II should be of
order\footnote{Here we have normalized the $BR$ with respect to
the $h\rightarrow b\bar{b}$ channel only, because the gauge boson
modes will be suppressed in the relevant FCNC region, Cf. Section
3.}
\begin{equation}
    BR^{II}(h\rightarrow t\,\bar{c})
    \sim\left(\frac{|V_{bc}|}{16\pi^2}\right)^2\,\alpha_{W}\
    G_F\,m_b^2\,\tan^2\beta\ \lambda^2_{H^+\,H^-\,h}\,,
    \label{estimateBR2HDM1}
\end{equation}
where $\lambda_{H^+\,H^-\,h}$ is defined here in units of $g$ and
dimensionless {as compared to Table~\ref{tab:trilineals}}. Clearly
a big enhancement factor $\tan^2\beta$ appears, but this does not
suffice. Fortunately, the trilinear couplings
$\lambda_{H^+\,H^-\,h}$ for $h=h^0,H^0$ {(but not for $h=A^0$)}
carry two additional sources of potential enhancement ({Cf.
Table~\ref{tab:trilineals}) which are absent in the MSSM case.
Take e.g. $h^0$, then we see that under appropriate conditions
(for example, large $\ta$ and large $\tb$) the trilinear coupling
behaves as $\lambda_{H^+\,H^-\,h^0}\sim
(m_{h^0}^2-m_{A^0}^2)\,\tan\beta/(M_W\,\mHp)$, and in this case
\begin{equation}
    BR^{II}(h^0\rightarrow t\,\bar{c})
    \sim\left(\frac{|V_{bc}|}{16\pi^2}\right)^2\,\alpha_{W}\
    G_F\,m_b^2\,\tan^4\beta\,
    \left(\frac{m_{A^0}^2-m_{h^0}^2}{M_W\,\mHp}\right)^2 \,.
    \label{estimateBR2HDM2}
\end{equation}
So finally $BR^{II}(h^0\rightarrow t\,\bar{c})$, and of course
$BR^{II}(h^0\rightarrow \bar{t}\,{c})$, can be augmented by a huge
factor $\tan^4\beta$ times the square of the relative splitting
among the CP-even Higgs decaying boson mass and the CP-odd Higgs
mass. Since the neutral Higgs bosons do not participate in the
loop form factors (Cf. Fig. 1 Ref.\,\cite{BejarNP1}), it is clear
that various scenarios can be envisaged where these mass
splittings can be relevant. In the next section this behaviour
will be borne out by explicit calculations showing that
$h^0\rightarrow t\,\bar{c}$ can be raised to the visible level in
the case of the Type II model. As for the Type I model the Higgs
trilinear coupling enhancement is the same, but in the charged
Higgs Yukawa coupling all quarks go with a factor $\cot\beta$;
hence when considering the leading terms in the loops that
contribute one sees that in the corresponding expression
(\ref{estimateBR2HDM1}) the term $m_b^2\,\tan^2\beta$ is traded
for $m_t\,m_c\,\cot^2\beta$, which is negligible at high
$\tan\beta$. Both sources of enhancement are needed, and this
feature is only tenable in the 2HDM II.  Of course one could
resort to the range $\tan\beta\ll 1$ for the Type I models, but
this is not theoretically appealing. {For example, for
$\tb\lesssim 0.1$  the top quark Yukawa coupling
$g_t=g\,m_t/(\sqrt{2}\,M_W\,\sin\beta)$, which is present in the
interaction Lagrangians above, is pushed into the non-perturbative
region $g_t^2/16\pi^2\gtrsim1$ and then our calculation would not
be justified. And what is worse: for the 2HDM I we would actually
need $\tb\leq {\cal O}(10^{-2})$ to get significant FCNC rates! }
In short, we consider that $BR^{I}(h\rightarrow
t\,\bar{c}+\,\bar{t}\,c)$ is essentially small (for all $h$), and
that these decays remain always invisible to speak of. Hereafter
we abandon the study of the decays (\ref{hFCNC}) for the 2HDM I
and restrict ourselves to the general 2HDM II.

\section{Numerical analysis}

Let us now substantiate the previous claims and provide the
precise numerical results of the full one-loop calculation of
$BR^{II}(h\rightarrow t\,\bar{c}+\bar{t}\,{c})$\,\footnote{Here
and throughout we use the notation $BR^{II}(h^0\rightarrow
t\,\bar{c}+\bar{t}\,{c})\equiv BR^{II}(h^0\rightarrow t\,\bar{c})+
BR^{II}(h^0\rightarrow \bar{t}\,{c})$.} as well as of the LHC
production rates of these FCNC events. We shall closely follow
the notation and methods of
Refs.\,\cite{GuaschNP1,BejarNP1,BejarRADCOR}. We refer the reader
to these references for more details. See also \cite{Hunter}  for
basic definitions in the general 2HDM framework  and
\cite{CGGJS,tb2HDM} for calculational techniques and further
illustration of the $\tan\beta$ enhancement in other relevant
Higgs processes both in the MSSM and the 2HDM. In what follows we
shall limit ourselves to present the final results of our
numerical analysis together with a detailed discussion,
interpretation and phenomenological application. We have
performed the calculations with the help of the numeric and
algebraic programs FeynArts, FormCalc and
LoopTools~\cite{FeynArts}. The calculation must obviously be
finite without renormalization, and indeed the cancellation of UV
divergences in the total amplitudes was verified explicitly.

The input set for our numerical analysis is given by the data row
\begin{equation}\label{row}
    (\mh,\mH,\mA,\mHp,\ta,\tb)
\end{equation}
made out of six independent parameters in the general 2HDM.
Remaining inputs as in \cite{PDB2000}. In practice there are some
phenomenological restrictions on the data (\ref{row}) which were
already described in \cite{BejarNP1} and references therein,
{particularly\,\cite{OPAL}}. It suffices to say here that the
Higgs boson masses in (\ref{row}) are much less restricted than
in the MSSM. Actually, the direct experimental searches put a
soft bound of $\sim 70\,GeV$ for the lightest neutral Higgs boson
and around $80\,GeV$ for the charged one $H^\pm$. These are much
more relaxed than the bound of $114.1\,GeV$ placed on the SM
Higgs boson\,\cite{LEPHWG}. In the MSSM, direct LEP 200 searches
set a limit of $\,\gtrsim 90\,GeV$ for $\mh$\,\cite{LEPHWG} and a
similar limit for $\mA$ which translates (approximately) to
$\mHp\simeq 120\,GeV$ due to the supersymmetric Higgs mass
relations. Moreover, in contrast to the general 2HDM, in the MSSM
case there is a theoretical upper bound $\mh\lesssim 122,
135\,GeV$ (for minimal and maximal top-squark mixing,
respectively) \,\cite{Hollik}. Concerning $\tan\beta$, in a
general 2HDM it is restricted only, by perturbativity arguments,
to lie within the approximate range
\begin{equation}\label{tbbound}
    0.1\lesssim\tan\beta\lesssim 60\,.
\end{equation}
{Strictly speaking the upper bound could be larger, but as
in\,\cite{BejarNP1} we shall stick to the above moderate range.
In practice, since Type I models are not considered, the effective
range for our calculation will be the high $\tb$ end of
(\ref{tbbound}).} For comparison, in the MSSM case there are
additional phenomenological restrictions that bring the lower
bound on $\tb$ to roughly $2.4$ for the so-called maximal $m_h^0$
scenario, and $10.5$ for the so-called no mixing
scenario\,\cite{LEPHWG}, the upper bound being the same.
Furthermore, in contrast to the MSSM, the parameter $\ta$ is free
in the 2HDM.
\begin{figure}[t]
    \centering
    \includegraphics[width=0.8\textwidth]{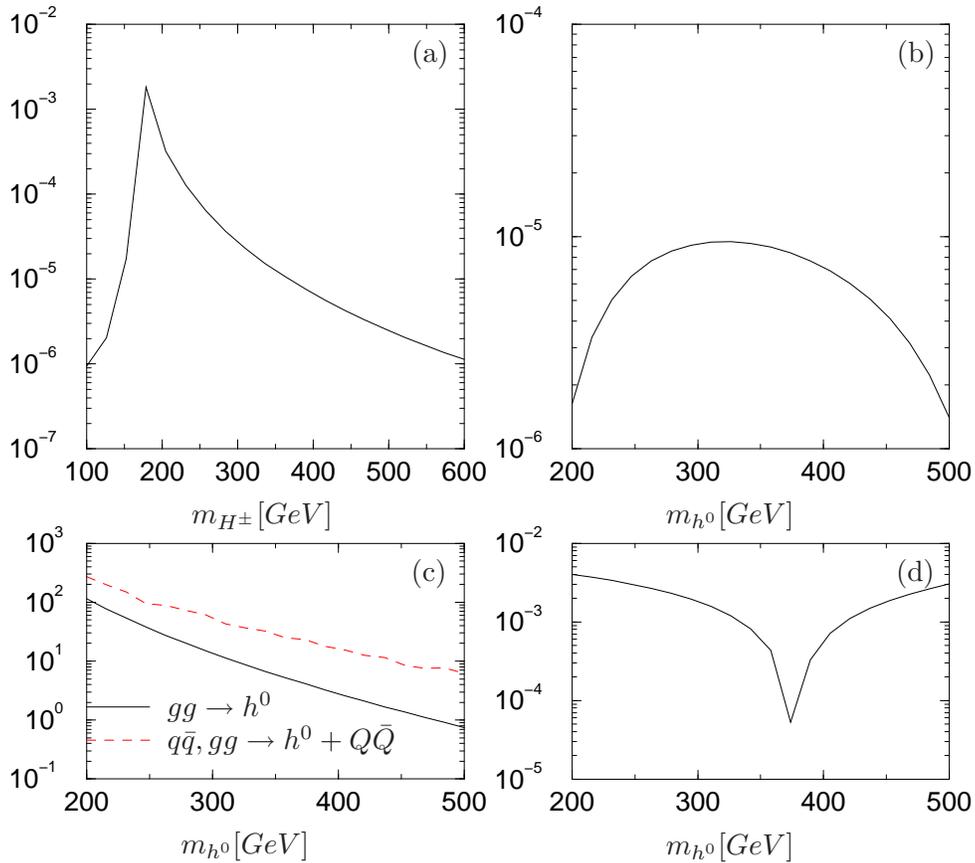}
    \caption{\textbf{(a)}$\, BR^{II}(h^0\rightarrow
    t\,\bar{c}+\bar{t}\,c)$
      versus $\mHp$; \textbf{(b)} Idem, versus $\mh$; \textbf{(c)} The
      production cross-section (in $pb$) of $h^0$ at the LHC versus its mass;
      \textbf{(d)} $\delta\rho^{\rm 2HDM}$ versus $\mh$, see the text. In these figures, when a parameter is not
      varied  it is  fixed as in eq.(\ref{inputparam}).}
      \label{fig:4figslh}
\end{figure}
\begin{figure}[htb]
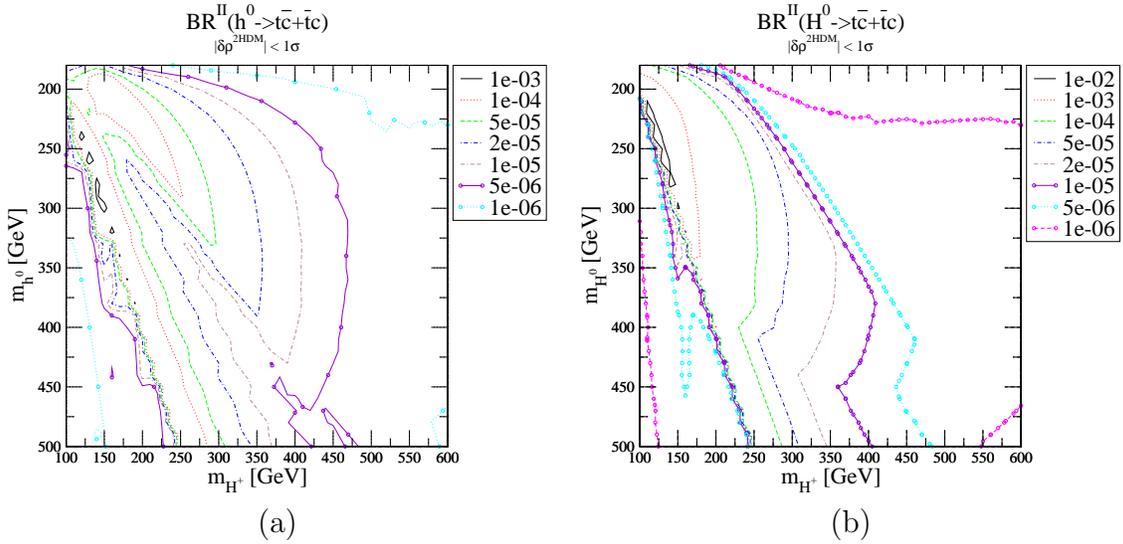

    \begin{tabular}{cc}
        \includegraphics*[width=0.45\textwidth]{br1} &
        \includegraphics*[width=0.45\textwidth]{brhh1} \\
        (a) & (b)
    \end{tabular}
    \caption{Contour lines
    in the $(\mHp,\mh)$-plane for the branching ratios (2HDM II case)  \textbf{(a)}
     $BR^{II}(h^0\rightarrow t\,\bar{c}+\bar{t}\,c$) and \textbf{(b)} $BR^{II}(H^0\rightarrow
      t\,\bar{c}+\bar{t}\,c$) assuming  $\delta\rho^{\rm 2HDM}$
      at $1\,\sigma$.}
    \label{fig:br1}
\end{figure}
\begin{figure}[t]
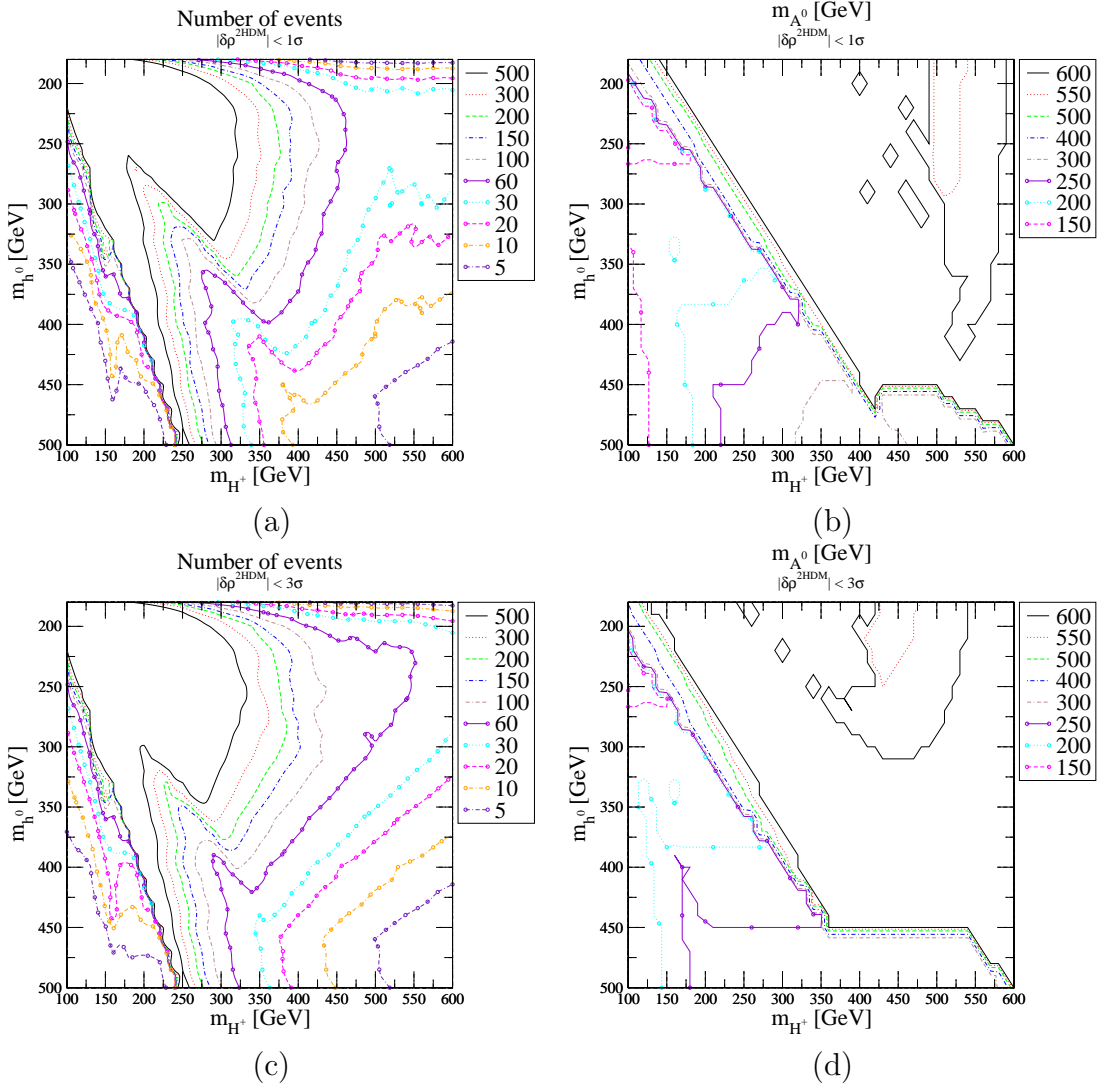

    \begin{tabular}{cc}
        \includegraphics*[width=0.44\textwidth]{number1} &
        \includegraphics*[width=0.44\textwidth]{ma1}\\
        (a) & (b) \\
         \includegraphics*[width=0.44\textwidth]{number3} &
        \includegraphics*[width=0.44\textwidth]{ma3} \\
        (c) & (d)
    \end{tabular}
    \caption{\textbf{(a)} Contour lines in the $(\mHp,\mh)$-plane for the
      maximum number of light CP-even Higgs FCNC events $h^0\rightarrow
      t\,\bar{c}+\bar{t}\,c$ produced at the LHC for
      $100\,fb^{-1}$ of integrated luminosity; \textbf{(b)}
{Contour lines showing the value of $\mA$ that maximizes the number of
events;}
\textbf{(c)}  As in (a) but within $\delta\rho^{\rm
    2HDM}$  at $3\,\sigma$; \textbf{(d)}  As in (b) but with
    $\delta\rho^{\rm 2HDM}$
        at $3\,\sigma$.}
    \label{fig:numberma1}
\end{figure}

\begin{figure}[t]
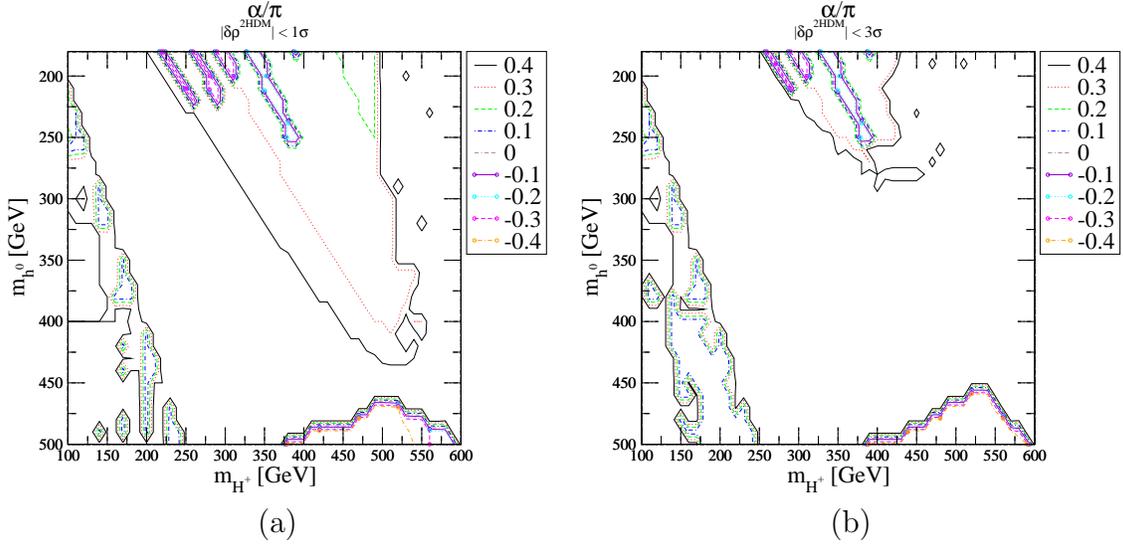

    \begin{tabular}{cc}
        \includegraphics*[width=0.45\textwidth]{a1} &
        \includegraphics*[width=0.45\textwidth]{a3}\\
        (a) & (b)
    \end{tabular}
    \caption{Contour lines $\alpha/\pi=const.$ ($\alpha$ is the mixing angle in the CP-even sector)
    corresponding to Fig. \ref{fig:numberma1} for $\delta\rho^{\rm 2HDM}$  at
      \textbf{(a)} $1\,\sigma$ and \textbf{(b)} $3\,\sigma$ .}
    \label{fig:a13}
\end{figure}
\begin{figure}[t]
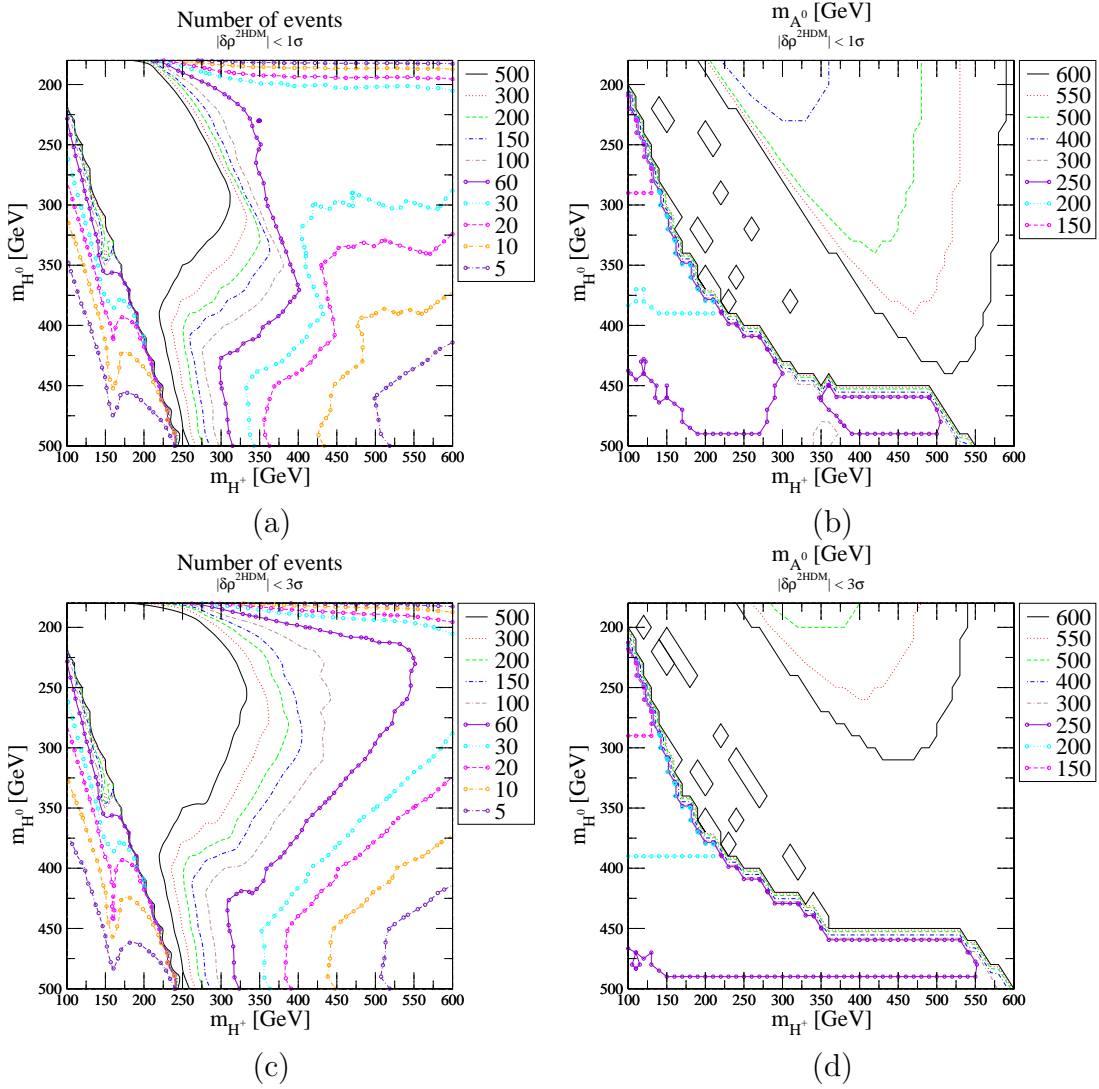

    \begin{tabular}{cc}
        \includegraphics*[width=0.44\textwidth]{numberhh1} &
        \includegraphics*[width=0.44\textwidth]{mahh1}\\
        (a) & (b) \\
        \includegraphics*[width=0.44\textwidth]{numberhh3} &
        \includegraphics*[width=0.44\textwidth]{mahh3} \\
        (c)&(d)
    \end{tabular}
    \caption{\textbf{(a)} Contour lines in the $(\mHp,\mH)$-plane for the
      maximum number of heavy CP-even Higgs FCNC events $H^0\rightarrow
      t\,\bar{c}+\bar{t}\,c$ (2HDM II case) produced at the LHC for
      $100\,fb^{-1}$ of integrated luminosity; \textbf{(b)} Corresponding contour lines for
      $\mA$;
      \textbf{(c)}  As in (a) but within $\delta\rho^{\rm 2HDM}$
        at $3\,\sigma$; \textbf{(d)}  As in (b) but with $\delta\rho^{\rm 2HDM}$
        at $3\,\sigma$.}
    \label{fig:numbermahh1}
\end{figure}

\begin{figure}[t]
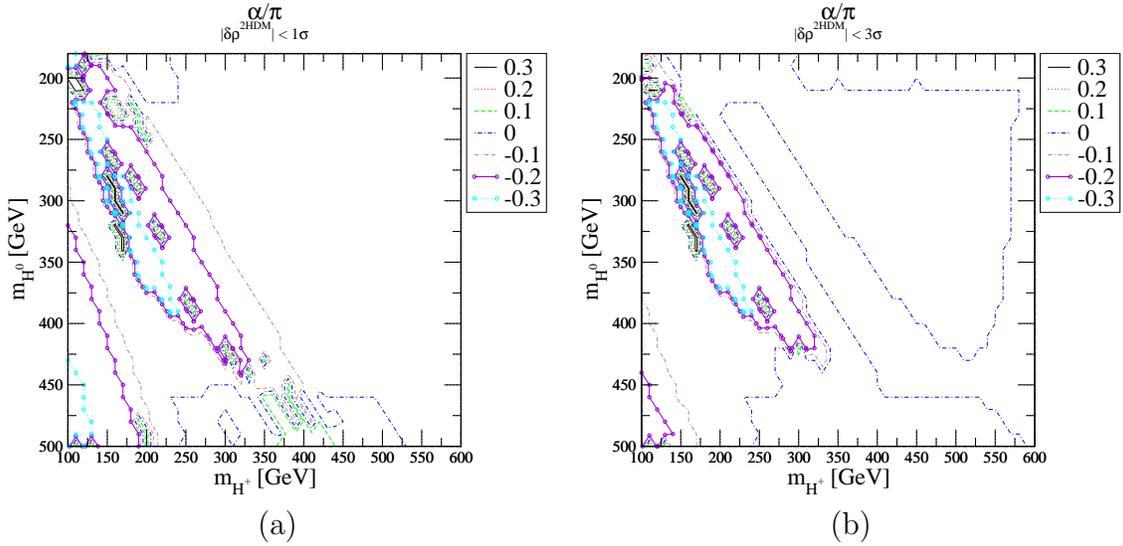

    \begin{tabular}{cc}
        \includegraphics*[width=0.45\textwidth]{ahh1}&
        \includegraphics*[width=0.45\textwidth]{ahh3}\\
        (a) & (b)
    \end{tabular}
    \caption{Contour lines $\alpha/\pi=const.$ as in Fig.\ref{fig:a13}, but for the heavy CP-even Higgs.}
    \label{fig:ahh13}
\end{figure}
Apart from the direct limit $\mHp\gtrsim 80\,GeV$ from LEP 200,
there are more stringent restrictions on the charged Higgs boson
mass. Although they are indirect and in this sense not so
indisputable, it is wise to assess their impact in our analysis.
These restrictions emerge from the virtual contributions to
$b\rightarrow s\,\gamma$ that have been computed at the NLO in
QCD\,\cite{Gambino}. From these calculations and from the
experimental limits on the radiative B-meson decays, these authors
can place a lower bound on the 2HDM II charged Higgs mass of $\mHp
\gtrsim 350\,GeV$\,\cite{Gambino}. We will, in general, apply this
lower bound, but since the direct and indirect searches complement
each other, we may disregard it in some parts of our numerical
analysis. The parts where we deviate will be clear in the text. We
recall that at the moment the MSSM may escape this indirect bound
because the positive charged Higgs virtual contributions to
$b\rightarrow s\,\gamma$ can be compensated for by negative stop
and chargino loops, if they are not too heavy. Therefore, in the
MSSM the charged Higgs can stay relatively light, $\mHp\gtrsim
120\,GeV$, just to comply with the aforementioned LEP 200 bounds
on $\mA$\,\cite{LEPHWG}.

Finally, we will restrict the input data row (\ref{row}) with the
important constraint from $\delta\rho$, extensively used in
\cite{GuaschNP1,BejarNP1} for the FCNC top quark decays in the
MSSM and the 2HDM. Namely, the one-loop corrections to the
$\rho$-parameter (i.e. the ratio between the charged and neutral
current Fermi constants) are bound within $|\delta\rho^{\rm 2HDM
}|\lesssim 0.1\%$. To be precise, the latter is the extra effect
that $\delta\rho$ can accommodate at one standard deviation
($1\,\sigma$) from the 2HDM fields beyond the SM contribution
\cite{GuaschNP1}. This is a stringent restriction that affects the
possible mass splittings among the Higgs fields of the 2HDM, and
its implementation in our codes does severely prevent the
possibility from playing with the Higgs boson masses to
artificially enhance the FCNC contributions.

With these restrictions in mind we have computed the number of
FCNC Higgs decay events into top quark at the LHC:
\begin{equation}\label{CS}
pp\to\ h +X\rightarrow t\,\bar{c}\,(\bar{t}\,c) +X \ \ \ \
(h=h^0,H^0,A^0)\,.
\end{equation}
The necessary cross-sections to compute the production of neutral
Higgs bosons at this collider, including {all known QCD
corrections}, have been computed by adapting the codes HIGLU {1.0}
and HQQ {1.0}\,\cite{Spira} -- originally written for the MSSM
case\,\cite{Spira2}-- to the general case of the
2HDM\footnote{{We have used the default parton distribution
functions and  renormalization/factorization scales used in these
programs, namely GRV94 with $\mu_R=\mu_F=m_h$  for HIGLU, and
CTEQ4L with {$\mu_R=\mu_F=\sqrt{\hat{s}}\equiv\sqrt{(p_h+p_Q
+p_{\bar Q})^2}$ } for  HQQ.}}. Folding the cross-sections with
the one-loop branching ratios of the processes (\ref{hFCNC}) we
have obtained the number of FCNC Higgs decay events at the LHC.
Let us first consider the branching ratios themselves. In
Fig.~\ref{fig:4figslh}a,b we show $BR^{II}(h^0\rightarrow
t\,\bar{c}+\bar{t}\,{c})$ for the lightest CP-even state. In
particular, Fig.~\ref{fig:4figslh}a shows $BR^{II}(h^0\rightarrow
t\,\bar{c}+\bar{t}\,c)$  versus the charged Higgs mass $\mHp$. In
this figure we fix the values of the parameters in (\ref{row})
which are not varying as follows:
\begin{equation}
\label{inputparam}
\begin{array}{l}
(\mh=350\,GeV,\mH=600\,GeV,\mA=550\,GeV,\mHp=375\,GeV,\\
\ta=30,\tb=60)
\end{array}
\end{equation}
After crossing a local maximum (associated to a pseudo-threshold
of the one-loop vertex function involving the $h^0\,H^+\,H^-$
coupling) the subsequently falling behavior of the $BR$ with
$\mHp$ clearly shows that the previously discussed bounds on
$\mHp$ are quite relevant. The branching ratio, however, stays
within $10^{-6}-10^{-5}$ for a wide range of heavy charged Higgs
masses extending up to $\mHp\leq 600\,GeV$ in
Fig.~\ref{fig:4figslh}a. Hence, for $\mHp$ heavy enough to
satisfy the indirect bounds from radiative $B$-meson
decays\,\cite{Gambino}, the maximum $BR$ is still sizeable. In
Fig.~\ref{fig:4figslh}c the production rate of $h^0$ bosons at the
LHC is shown as a function of $\mh$, for fixed parameters
(\ref{row}). The production cross-sections for the subprocesses
\begin{equation}\label{CS2}
gg\to\ h^0 +X\,, \ \ \ \ \ gg,qq\to\ h^0 +Q\bar{Q}\,,
\end{equation}
contributing to (\ref{CS}) in the case of the light CP-even Higgs
$h^0$ are explicitly separated in Fig.~\ref{fig:4figslh}c. The
gluon-gluon fusion process proceeds at one-loop and the
{$h^0 Q\bar{Q}$ associated production proceeds at tree-level\,\cite{Spira3}.}
Similar subprocesses and results apply for $H^0$ and $A^0$
production. {At large $\tb$ and the larger the Higgs
boson masses the particular associated production mechanism with
the bottom quark, $Q=b$, i.e. $h^0\ b\bar{b}$, becomes dominant by
far}. All other mechanisms for Higgs boson production in Type II
models\,\cite{Spira2,Spira3,CoarasaJS}, like vector-boson fusion
{(which contributes also to $h^0 Q\bar{Q}$ when $Q$ are light quarks)},
{vector-boson bremsstrahlung ($q\bar{q}\to h\,V$)} and associated $t\,\bar{t}$
production, are subdominant at large $\tb$ and can be neglected
for our purposes. Admittedly, some of these mechanisms can be
relevant for Higgs boson production in the case of the Type I
2HDM at low $\tb$, but we have already warned that the
corresponding FCNC branching ratios are never sufficiently high.

The control over $\delta\rho^{\rm 2HDM}$ is displayed in
Fig.~\ref{fig:4figslh}d. Recall that $\delta\rho$ is not sensitive
to the mass splitting between $\mh$ and $\mH$, because of
$CP$-conservation in the gauge boson sector, but it does feel all
the other mass splittings among Higgs bosons, charged and neutral.
A more systematic search of $BR$ values in the parameter space is
presented in Figs. \ref{fig:br1}a,b corresponding to
$BR^{II}(h^0\rightarrow t\,\bar{c}+\bar{t}\,c)$ and
$BR^{II}(H^0\rightarrow t\,\bar{c}+\bar{t}\,c)$ respectively. Here
we have scanned independently on the parameters (\ref{row}) while
holding the $\delta\rho^{\rm 2HDM}$ bound at $1\,\sigma$. The
contour lines in these figures represent the locus of points in
the $(\mHp,\mh)$-plane giving maximized values of the $BR$ in the
2HDM II. Let us remark that the highest value of $\tb$ is always
preferred, and therefore all these contour lines correspond to
$\tb=60$.

In practice, to better assess the possibility of detection at the
LHC, one has to study the production rates of the FCNC events.
These are determined by combining the production cross-sections of
neutral 2HDM II Higgs bosons at the LHC and the FCNC branching
ratios. If we just adopt the mild LEP bound $\mHp\gtrsim 80\,GeV$
and let $\mHp$ approach the maximum in Fig. \ref{fig:4figslh}a
then the $BR$ can be as large as $10^{-3}$ and the number of FCNC
events can be huge, at the level of ten thousand per
$100\,fb^{-1}$ of integrated luminosity. But of course the region
near the maximum is too special. Moreover, if we switch on the
above mentioned indirect bound from $b\rightarrow
s\,\gamma$\,\cite{Gambino}, then the typical $BR$ is much smaller
(of order $10^{-5}$) and the number of events is reduced
dramatically, at a level of hundred or less for the same
integrated luminosity. On the other hand it may well happen that
there are regions of parameter space where $BR\sim 10^{-5}$ (see
Fig. \ref{fig:br1}) but the production cross-section is too small
because the decaying Higgs boson is too heavy. Therefore, it is
the product of the two quantities that matters.

The systematic search of the regions of parameter space with the
maximum number of FCNC events for the light CP-even Higgs is
presented in the form of contour lines in the multiple
Fig.~\ref{fig:numberma1}. For instance, each isoline in
Figs.~\ref{fig:numberma1}\,a,c corresponds to a fixed number of
produced FCNC events at the LHC while keeping the value of
$\delta\rho^{\rm 2HDM}$ within $1\,\sigma$ or $3\,\sigma$
respectively of its central experimental value. When scanning over
the parameter space (\ref{row}) we have found again that
$\tan\beta$ is preferred at the highest allowed value
($\tan\beta=60$) -- for Type II models. We have also determined
(see Figs.\,\ref{fig:numberma1}\,b,d) the corresponding contour
lines for $\mA$ associated to these events. The $\mA$-lines are
important because the FCNC processes under consideration are
sensitive to the mass splittings between $\mA$ and the
corresponding decaying Higgs boson, see e.g.
eq.(\ref{estimateBR2HDM2}) {and Table~\ref{tab:trilineals}}. The
combined figures \ref{fig:numberma1}\,a-d are very useful because
they give a panoramic view of the origin of our results in the
parameter space. To complete the map of the numerical analysis we
provide Fig. \ref{fig:a13} in which we have projected the contour
lines of the CP-even mixing angle $\alpha$ associated to the
previous plots. For a given contour line $\alpha/\pi=const.$, the
set of inner points have a value of $\alpha/\pi$ smaller than the
one defined by the line itself. In particular, the large domains
in Figs. \ref{fig:a13}a,b without contour lines correspond to
$\alpha/\pi>0.4$ and so to relatively large (and positive)
$\tan\alpha$.  There are a few and small neighborhoods where the
FCNC rates for $h^0$ can be sizeable also for small $\ta$.

Knowing that high $\tan\alpha$ is generally preferred by
$h^0\rightarrow t\,\bar{c}+\bar{t}\,c$, and noting from
Fig.\,\ref{fig:numberma1} that large mass splittings between $\mh$
and $\mA$ are allowed, we find that the trilinear coupling
$\lambda_{H^+\,H^-\,h^0}$ can take the form
$\lambda_{H^+\,H^-\,h^0}\sim
(m_{h^0}^2-m_{A^0}^2)\,\tan\beta/(M_W\,\mHp)$. Hence it provides a
substantial additional enhancement beyond the $\tan\beta$ factor.
One can check from the approximate formula (\ref{estimateBR2HDM2})
that the maximum FCNC branching ratios $BR^{II}(h^0\rightarrow
t\,\bar{c}+\bar{t}\,c)$ can eventually reach the $10^{-5}$ level
even in regions where the charged Higgs boson mass preserve the
stringent indirect bounds from radiative $B$-meson
decays\,\cite{Gambino}. These expectations are well in agreement
with the exact numerical analysis presented in
Fig.\,\ref{fig:br1}, thus showing that eq.(\ref{estimateBR2HDM2})
provides a reasonable estimate, and therefore a plausible
explanation for the origin of the maximum contributions. {As a
matter of fact, we have checked that the single (finite)
  Feynman diagram giving rise to the estimation~(\ref{estimateBR2HDM2})
  -- the {one-loop vertex Feynman diagram} with a couple of charged Higgs bosons
  and a bottom quark in the loop -- reproduces the full result with an
  accuracy better than 10\% for $\tb\gtrsim 10-20$. At lower $\tb$ values
  large deviations are possible but, as warned before,
  eq.~(\ref{estimateBR2HDM2}) is expected to be valid only at large
  $\tb$. {Furthermore}, for low values of $\tb\lesssim 20$ the FCNC $BR$s are
  too small to be of any phenomenological interest.} The
exact numerical analysis is of course based on the full expression
for the branching ratio
\begin{equation}
    BR^{II}(h^0\rightarrow t\,\bar{c}+\bar{t}\,c)=\frac{\Gamma
      (h^0\rightarrow t\,\bar{c}+\bar{t}\,c)}{\Gamma (h^0\rightarrow
      b\,\bar{b})+\Gamma (h^0\rightarrow
      t\,\bar{t})+\Gamma(h^0\rightarrow V\,V)+\Gamma(h^0\rightarrow H\,H)}\,, \label{fullBRh}
\end{equation}
where all decay widths in the denominator of this formula have
been computed at the tree-level in the 2HDM II, since this
provides a consistent description of eq.~(\ref{fullBRh})
at leading order. Here we have defined
\begin{equation}\label{totalw1}
\Gamma(h^0\rightarrow V\,V)\equiv \Gamma(h^0\rightarrow
W^+W^-)+\Gamma(H^0\rightarrow Z\,Z)\,,
\end{equation}
\begin{equation}\label{totalw2}
\Gamma(h^0\rightarrow H\,H)\equiv \Gamma(h^0\rightarrow
A^0\,A^0)+\Gamma(h^0\rightarrow H^+\,H^-)\,.
\end{equation}
{We disregard the loop induced decay channels, since they have
  branching ratios below the percent level all over the parameter
  space. The $\tau$-lepton decay channel is also neglected, since it is
  suppressed by a factor of ${\cal O}(10^{-2})$ with respect the $b\bar{b}$-channel in the whole 2HDM
  parameter space.}
In general the effect of the gauge boson channels $h^0\rightarrow
W^+W^-, ZZ$ in the $BR$ (\ref{fullBRh}) is not so important as in
the SM, actually for $\beta=\alpha$ they vanish in the $h^0$ case
because they are proportional to $\sin^2(\beta-\alpha)$. This is
approximately the case for large $\ta$ and large $\tb$, the
dominant FCNC region for $h^0$ decay (Cf. Fig.\ref{fig:a13}a and
\ref{fig:a13}b).  In this region, the mode $h^0\rightarrow
t\,\bar{t}$ is, when kinematically allowed, suppressed:
$BR(h^0\rightarrow
t\,\bar{t})\propto\cos^2\alpha/\sin^2\beta\rightarrow 0$ (Cf.
Eq.\,(\ref{Hff})). On the other hand there are domains in our
plots where the decays $h^0\rightarrow H^+\,H^-$ and
$h^0\rightarrow A^0\,A^0$ are kinematically possible and
non-(dynamically) suppressed. {Indeed, this can be checked from
the explicit structure of the trilinear couplings $h^0\,H^+\,H^-$
and $h^0\,A^0\,A^0$ in Table~\ref{tab:trilineals}; in the dominant
region for the decays $h^0\rightarrow t\,\bar{c}+\bar{t}\,{c}$
both of these couplings are $\tb$-enhanced}. {Nevertheless the
decay $h^0\rightarrow A^0\,A^0\,$ is only possible for}
$\mA<\mh/2\,$, and since the optimal FCNC regions demand the
largest possible values of $\mA$, this decay is kinematically
blocked there. On the other hand the mode $h^0\rightarrow
H^+\,H^-$ is of course allowed if we just take the aforementioned
direct limits on the 2HDM Higgs boson masses. But it is never
available if we apply the indirect bound from $b\rightarrow
s\,\gamma$ on the charged Higgs mass mentioned above, unless
$\mh>2\mHp>700\,GeV$, in which case $h^0$ is so heavy that its
production cross-section is too small for FCNC studies to be
further pursued.

The corresponding results for the heavy CP-even Higgs boson are
displayed in Figs. \ref{fig:numbermahh1} and \ref{fig:ahh13}. The
exact formula for the $BR$ in this case reads
\begin{equation}
    BR^{II}(H^0\rightarrow t\,\bar{c}+\bar{t}\,c)=\frac{\Gamma
      (H^0\rightarrow t\,\bar{c}+\bar{t}\,c)}{\Gamma (H^0\rightarrow
      b\,\bar{b})+\Gamma (H^0\rightarrow
      t\,\bar{t})+\Gamma(H^0\rightarrow V\,V)+\Gamma(H^0\rightarrow H\,H)}\,,
      \label{fullBRH}
\end{equation}
where we have defined
\begin{equation}\label{totalw3}
\Gamma(H^0\rightarrow V\,V)\equiv \Gamma(H^0\rightarrow
W^+W^-)+\Gamma(H^0\rightarrow Z\,Z)\,,
\end{equation}
\begin{equation}\label{totalw4}
\Gamma(H^0\rightarrow H\,H)\equiv \Gamma(H^0\rightarrow
h^0\,h^0)+\Gamma(H^0\rightarrow A^0\,A^0)+\Gamma(H^0\rightarrow
H^+\,H^-)\,.
\end{equation}
From the contour lines in Fig.\,\ref{fig:numbermahh1}a,c it is
patent that the number of FCNC top quark events stemming from
$H^0$ decays is comparable to the case of the lightest Higgs
boson. However, Fig. \ref{fig:ahh13}a,b clearly reveals that
these events are localized in regions of the parameter space
generally different from the $h^0$ case, namely they prefer
$\ta\simeq 0$. Even so, there are some ``islands'' of events at
large $\ta$. This situation is complementary to the one observed
for $h^0$ in Fig.\,\ref{fig:a13}. However, in both cases
these isolated regions are mainly concentrated in the segment
$\mHp< 350\,GeV$. Therefore, if the bound on $\mHp$ from
$b\rightarrow s\,\gamma$ is strictly preserved, it is difficult to
find regions of parameter space where the two CP-even states of a
general 2HDM II may both undergo a FCNC decay of the type
(\ref{hFCNC}).

In the dominant regions of the FCNC mode $H^0\rightarrow
t\,\bar{c}$ (where $\ta$ is small and $\tb$ is large), the decay
of $H^0$ into the $t\bar{t}$ final state is suppressed:
$BR(H\rightarrow
t\,\bar{t})\propto\sin^2\alpha/\sin^2\beta\rightarrow 0$. In the
same regions the gauge boson channels in (\ref{fullBRH}) are
{suppressed too because} $\Gamma(H^0\rightarrow W^+W^-, ZZ)\propto
\cos^2(\beta-\alpha)$. In principle the heavy CP-even Higgs boson
$H^0$ also could (as $h^0$) decay into $A^0\,A^0$ and $H^+\,H^-$.
But there is a novelty here with respect to the $h^0$ decays, in
that there could be regions where  $H^0$ could decay into the
final state $h^0\,h^0$. This contingency has been included
explicitly in eq.(\ref{totalw4}). However, in practice, neither
one of these three last channels is relevant in the optimal FCNC
domains of parameter space. {First, the decay $H^0\rightarrow
h^0\,h^0$, although it is kinematically possible, is dynamically
suppressed in the main FCNC region for $H^0$. This can be seen
from Table~\ref{tab:trilineals}, where the trilinear coupling
$H^0\,h^0\,h^0$ becomes vanishingly small at large $\tb$ and
small $\ta$. Second, the coupling ${H^0A^0A^0}$ in
Table~\ref{tab:trilineals} is non-suppressed in the present
region, but again the mode $H^0\rightarrow A^0\,A^0$ is
kinematically forbidden in the optimal FCNC domains because the
latter favor large values of the CP-odd mass (see
Fig.\,\ref{fig:numbermahh1}b,d). Third, although in these domains
the decay $H^0\rightarrow H^+\,H^-$ is also non-dynamically
suppressed (see the corresponding trilinear coupling in
Table~\ref{tab:trilineals}), it becomes kinematically shifted to
the high mass range $\mH> 700\,GeV$ if we switch on the indirect
bound from $b\rightarrow s\,\gamma$. Obviously, in  this latter
case the $H^0$ production cross section becomes too small and the
FCNC study has no interest.} All in all the contributions from
(\ref{totalw1}),(\ref{totalw2}),(\ref{totalw3}) and
(\ref{totalw4}) are irrelevant for $\mh,\mH<700\,GeV$ as their
numerical impact on $B^{II}(h^0,H^0\rightarrow
t\,\bar{c}+\bar{t}\,c)$ is negligible. Our formulae
(\ref{fullBRh}) and (\ref{fullBRH}) do contain all the decay
channels and we have verified explicitly these features.

As remarked before, in general the most favorable regions of
parameter space for the FCNC decays of $h^0$ and $H^0$  do not
overlap much. The trilinear Higgs boson self-couplings {in
Table~\ref{tab:trilineals}} (also the fermionic ones) are
interchanged when performing the simultaneous substitutions
$\alpha\rightarrow\pi/2-\alpha$ and $\mh\rightarrow\mH$
\,\cite{BejarNP1}. Furthermore, the LHC production rates of the
neutral Higgs bosons fall quite fast with the masses of these
particles, as seen e.g. in Fig.~\ref{fig:4figslh}b for the $h^0$
state. As a consequence that exchange symmetry on the branching
ratios does not {go over} to the final event rates, so in practice
the number of FCNC events from $H^0$ decays are smaller (for the
same values of the other parameters) as compared to those for
$h^0$; thus $H^0$ requires e.g. lighter charged Higgs masses to
achieve the same number of FCNC events as $h^0$. {As for the
CP-odd state $A^0$, we have seen that it plays an important
indirect dynamical role on the other decays through the trilinear
couplings in Table~\ref{tab:trilineals}, but its own FCNC decay
rates never get a sufficient degree of enhancement due to the
absence of the relevant trilinear couplings, so we may discard it
from our analysis.}

We notice that this picture is consistent with the decoupling
limit in the 2HDM: for
  $\alpha\to\beta$, the heaviest  CP-even Higgs boson ($H^0$)
  behaves as the SM Higgs boson, whereas $h^0$ decouples from the electroweak gauge bosons
  and may develop enhanced
  couplings to up and down-like quarks, depending on whether $\tan\beta$ is small or
  large respectively;
  in the opposite limit   ($\alpha\to\beta-\pi/2$), it is $h^0$ that behaves as
  $H^{SM}$, while $H^0$
  decouples from gauge bosons and may develop the same enhanced couplings to quarks as
  $h^0$ did in the previous case.
  Indeed these are the situations that we
  find concerning the FCNC decay rates.
We recall that the numerical results presented in our figures
correspond to an integrated luminosity of $100\,fb^{-1}$.
{However, the combined ATLAS and CMS detectors might eventually
accumulate a few hundred inverse femtobarn\,\cite{ATLAS,CMS}.
Therefore, hopefully, a few hundred FCNC events (\ref{hFCNC})
could eventually be collected in the most optimistic scenario}.
Actually, the extreme rareness of these events in the SM suggests
that if only a few of them could be clearly disentangled, it
should suffice to claim physics beyond the SM.

\section{Discussion and conclusions}

Detection strategies at the CERN-LHC collider for the search of
the SM Higgs boson, and also for the three spinless fields of the
MSSM Higgs sector, have been described in great detail in many
places of the
literature\,\cite{Hunter,ATLAS,CMS,Gianotti,CarenaHaber}, but not
so well for the corresponding charged and neutral Higgs bosons of
the general 2HDM. The result is that the discovery of the SM
Higgs boson is guaranteed at the LHC in the whole presumed range
$100\,GeV\lesssim m_H\lesssim 1\,TeV$. However, the discovery
channels are different in each kinematical region and sometimes
the most obvious ones are rendered useless. For example, due to
the huge irreducible QCD background from $b\,\bar{b}$ dijets, the
decay mode $H^{SM}\rightarrow b\,\bar{b}$ is difficult and one
has to complement the search with many other channels,
particularly
$H^{SM}\rightarrow\gamma\,\gamma$\,\,\cite{ATLAS,CMS}. {We have
shown in this work that there are scenarios in the 2HDM parameter
space where alternative decays, like the FCNC modes
$h^0\rightarrow t\,\bar{c}+\bar{t}\,c$ and $H^0\rightarrow
t\,\bar{c}+\bar{t}\,c$, can also be useful}. For instance, in the
$h^0$ case, this situation occurs when $\tan\beta$ and
$\tan\alpha$ are both large and the CP-odd state is much heavier
than the CP-even ones. The potential enhancement is then
spectacular and it may reach up to ten billion times the SM value
$BR(H^{SM}\rightarrow t\,\bar{c})\sim 10^{-15}$, thereby bringing
the maximum value of the FCNC branching ratio $BR(h^{0}\rightarrow
t\,\bar{c})$ to the level of $\sim 10^{-5}$. As a matter of fact,
the enhancement would be much larger were it not because we
eventually apply the severe (indirect) lower bound  on the
charged Higgs mass from $b\rightarrow s\,\gamma$\,\cite{Gambino}.
Although these decays have maximal ratios below
$BR(h\rightarrow\gamma\gamma)\sim 10^{-3}$, they should be
essentially free of QCD background \footnote{Misidentification of
$b$-quarks as
 $c$-quarks in $tb$ production might be a source of background to
 our FCNC events. However, to rate the actual impact of that
 misidentification one would need a dedicated simulation of the
 signal versus background, which is beyond the scope of this
 paper.}.

While in the MSSM almost the full $(m_{A^0},\tan\beta)$-parameter
space is covered, with better efficiency at high $\tan\beta$
though, we should insist that within the general 2HDM the tagging
strategies are not so well studied and one would like to have
further information to disentangle the MSSM scenarios from the
2HDM ones. Here again the study of the FCNC Higgs decays can play
a role. Of course the statistics for the FCNC Higgs decays is poor
due to the weakness of the couplings and the large masses of the
Higgs bosons to be produced. However, in the favorable regions,
which are generally characterized by large values of $\tan\beta$
and of $\tan\alpha$, one may collect a few hundred events of the
type (\ref{hFCNC})-- mainly from $h^0$ -- in the high luminosity
phase of the LHC.  As we have said, this is basically due to the
enormous enhancement that may undergo the FCNC decay rates, but
also because in the same regions of parameter space where the
$BR$'s are enhanced, also the LHC production rates of the Higgs
bosons can be significantly larger (one order of magnitude) in
the 2HDM II as compared to the SM.

Interestingly enough, in many cases one can easily distinguish
whether the enhanced FCNC events (\ref{hFCNC}) stem from the
dynamics of a general, unrestricted, 2HDM model, or rather from
some supersymmetric mechanisms within the MSSM. This is already
obvious from the fact that the ranges of neutral and charged
Higgs boson masses in the 2HDM case can be totally incompatible
with the corresponding ones in the MSSM. But there are many other
ways to discriminate these rare events. For instance, in the 2HDM
case the CP-odd modes $A^0\rightarrow t\,\bar{c}+\bar{t}\,c$ are
completely hopeless whereas in the MSSM they can be
enhanced\,\cite{GuaschNP1,Curiel, Demir,unpublished}. Using this
information in combination {with} the masses of potentially
detected Higgs bosons could be extremely useful to pinpoint the
supersymmetric or non-supersymmetric nature of them. We may
describe a few specific strategies. As it was first shown in
Ref.\,\cite{GuaschNP1}, the leading SUSY-FCNC effects associated
to the $h\,t\,c$ vertices ($h=h^0\,,H^0\,,A^0$) come from the
FCNC gluino interactions which are induced by potentially large
misalignments of the quark and squark mass
matrices\,\cite{Duncan}. These effects are not particularly
sensitive to $\tan\beta$ and they can be very sizeable for both
high and moderately low values of this parameter. This sole fact
can be another distinguishing feature between FCNC events
(\ref{hFCNC}) of MSSM or 2HDM origin. If, for example, a few of
these events were observed and at the same time the best MSSM
fits to the electroweak precision data would favor moderate
values of $\tan\beta$, say in the range $10-20$, then it is clear
that those events could originate in the FCNC gluino interactions
but in no way within the context of the general 2HDM. In this
respect it should be mentioned that the  FCNC gluino couplings
recently became more restricted from the low-energy meson data
\,\cite{Besmer}, and will presumably become further restricted in
the near future. The reason being that the same couplings are
related, via $SU(2)$ gauge invariance and CKM rotation, to those
affecting the down-like quark sector, which will most likely
become constrained by the increasingly more precise low-energy
meson physics\,\cite{Besmer,Pokorsky}. In that circumstance the
only source of FCNC Higgs decays in the MSSM will stem purely
from the electroweak interactions within the super-CKM basis.
Then, in the absence of these SUSY-QCD FCNC effects, we could
judiciously conclude from the work of Ref.\cite{GuaschNP1} -- in
which both the SUSY-QCD \textit{and} the SUSY electroweak
contributions were computed for the $h\,t\,c$ vertices -- that
the FCNC rates in the MSSM should diminish dramatically (two to
three orders of magnitude). In such case we can imagine the
following ``provocative'' scenario. Suppose that the LHC finds a
light neutral Higgs boson of mass $\lesssim 140\,GeV$
(suggestively enough, in a mass range near the MSSM upper bound
for $\mh$!) and subsequently, or about simultaneously, a charged
Higgs boson and another neutral Higgs boson both with {masses
around $400\,GeV$ or more}. At this point one could naively
suspect that a MSSM picture out these findings is getting somehow
confirmed. If, however, later on a few FCNC events (\ref{hFCNC})
are reported and potentially ascribed to the previously
discovered heavy neutral Higgs boson {(presumably $H^0$)}, then
the overall situation could not correspond at all to the MSSM,
while it could be perfectly compatible with the 2HDM II.
Alternatively, suppose that the FCNC gluino couplings were not
yet sufficiently restricted, but (still following the remaining
hypotheses of the previous example) a {third} neutral Higgs boson
(presumably $A^0$) is found, also accompanied with a few FCNC
events. Then this situation would be incompatible with the 2HDM
II, and in actual fact it would put forward strong (indirect)
evidence of the MSSM!!\, \

We should also mention that there are other FCNC Higgs decay
modes, as for example $h\rightarrow b\,\bar{s}+\bar{b}\,s$, which
could be, in principle, competitive with the top quark modes
(\ref{hFCNC}). In some cases these bottom modes can be highly
enhanced in the MSSM case\,\cite{Curiel,Demir}.  Actually, a more
complete assessment of the FCNC bottom modes in the MSSM case --
namely one which takes {also} into account the supersymmetric
contributions to the highly restrictive radiative B-meson decays
-- shows that they are eventually rendered at a similar level of
the top modes under study in most of the parameter
space\,\cite{unpublished}.

To summarize, the FCNC decays of the Higgs bosons into top quark
final states can be a helpful complementary strategy to search for
signals of physics beyond the SM in the LHC. Our
{comprehensive} numerical analysis shows that the FCNC studies are
feasible for CP-even Higgs masses up to about $500\,GeV$. While
the statistics of these FCNC decays is of course poor, the
advantage is that a few tagged and well discriminated events of
this sort could not be attributed by any means to the SM, and
therefore should call for various kinds of new physics. In this
paper we have shown that a general 2HDM II is potentially
competitive to be ultimately responsible for these FCNC decays,
if they are ever found, and we have exemplified how to
discriminate this possibility from the more restricted one
associated to the MSSM.


{\bf Acknowledgments}

J.G. and J.S. are thankful to M. Spira for fruitful discussions.
This collaboration is part of the network ``Physics at
Colliders'' of the European Union under contract
HPRN-CT-2000-00149. The work of S.B. has been supported in part
by CICYT under project No. FPA2002-00648, and that of J.S.  by
MECYT and FEDER under project FPA2001-3598.


\begin{thebibliography}{99}




\bibitem{GIM}  S.L. Glashow, J. Iliopoulos, L. Maiani , \textit{Phys. Rev.}
    \textbf{D 2} (1970) 1285.

\bibitem{LorenzoD} J.L.\,Diaz-Cruz, R. Martinez, M.A. Perez, A.
Rosado, \textit{Phys. Rev.} \textbf{D 41} (1990) 891.

\bibitem{Eilam:1991zc}  G. Eilam, J.L. \,Hewett, A.
    Soni, \textit{Phys. Rev.} \textbf{D 44} (1991) 1473.

\bibitem{JASaavedra03} J.A.\,Aguilar-Saavedra, B.M. Nobre, \textit{Phys. Lett. }\textbf{B 553}
    (2003) 251, \texttt{hep-ph/0210360}.

\bibitem{Mele}  B.~Mele, S.~Petrarca, A.~Soddu, \textit{Phys. Lett.} \textbf{%
      B 435} (1998) 401, \texttt{hep-ph/9805498};
G.~Eilam, J.L.\,Hewett, A. Soni, \textit{Phys.
      Rev.} \textbf{D 59} (1998) 039901, Erratum.

\bibitem{Divitiis}  G.M. de Divitiis, R.~Petronzio, L. Silverstini, \textit{%
      Nucl. Phys. } \textbf{B 504} (1997) 45, \texttt{hep-ph/9704244}.

\bibitem{MSSMreps}
    H.P.~Nilles, \textit{Phys. Rep. } \textbf{110} (1984) 1;
H.E.~Haber, G.L.~Kane,  \textit{Phys. Rep. } \textbf{117} (1985)  75.

\bibitem{GuaschNP1}  J. Guasch, J. Sol{{\`a}}, \textit{Nucl. Phys. }\textbf{B 562}
    (1999) 3, \texttt{hep-ph/9906268}.

\bibitem{Yuan}  J.L.\,Diaz-Cruz, Hong-Jian He, C.-P. Yuan \textit{Phys. Lett. }\textbf{B 530}
    (2002) 179, \texttt{hep-ph/0103178}.


\bibitem{BejarNP1}  S. B{\'e}jar, J. Guasch, J. Sol{{\`a}}, \textit{%
      Nucl. Phys. } \textbf{B 600} (2001) 21, \texttt{hep-ph/0011091}.

\bibitem{Hunter}  J.F.\,Gunion, H.E.\,Haber, G.L.\,Kane, S. Dawson, \textit{The
      Higgs Hunters' Guide} (Addison-Wesley, Menlo-Park, 1990).

\bibitem{BejarRADCOR}  S. B{\'e}jar, J. Guasch, J. Sol{{\`a}}, in: Proc.
    of the 5th International Symposium on Radiative Corrections
    (RADCOR 2000), Carmel, California, 11-15 Sep 2000, \texttt{hep-ph/0101294}.

\bibitem{Saavedra}  J.A.\,Aguilar-Saavedra,  G.C. Branco,
  \textit{Phys. Lett. }\textbf{B 495} (2000) 347, \texttt{hep-ph/0004190}.


\bibitem{Hou} W.S. Hou, \textit{%
      Phys. Lett. }\textbf{B 296} (1992) 179.

\bibitem{Curiel} A.M.\,Curiel, M.J.\,Herrero, D.\,Temes,
     \textit{Phys. Rev. }\textbf{D 67} (2003) 075008, \texttt{hep-ph/0210335}.


 \bibitem{Demir}    D. A. Demir, \textit{Higgs couplings to quarks with supersymmetric CP anf
 flavor violations}, \texttt{hep-ph/0303249}.

 \bibitem{Brignole} A. Brignoli, A. Rossi, \textit{Lepton flavor violating decays of
 supersymmetric Higgs bosons}, \texttt{hep-ph/0304081}.

\bibitem{unpublished} {S. B{\'e}jar, F. Dilm{\'e}, J. Guasch, J. Sol{\`a}, in
preparation.}


\bibitem{CGGJS}  J.A. Coarasa, D. Garcia, J. Guasch, R.A. Jim{{\'e}}nez, J.
    Sol{{\`a}}, \textit{Eur. Phys. J.} \textbf{C 2} (1998) 373,
    \texttt{hep-ph/9607485}.


\bibitem{tb2HDM}  J.A. Coarasa, J. Guasch, W. Hollik  J. Sol{\`a},
    \textit{Phys. Lett.} \textbf{ B 442} (1998) 326, \texttt{hep-ph/9808278}.

\bibitem{FeynArts}  FeynArts: J. K{\"u}blbeck, M. B{\"o}hm, A. Denner,
    \textit{Comp. Phys. Comm}. \textbf{60} (1990) 165;
FormCalc and  LoopTools: T. Hahn, M. P{\'e}rez-Victoria, \textit{Comp. Phys.
      Comm}. \textbf{118} (1999) 153, \texttt{hep-ph/9807565};
{G.~J.~van Oldenborgh,
Comput.\ Phys.\ Commun.\  {\bf 66} (1991) 1;
J.~A.~Vermaseren, \textit{New features of FORM}, \texttt{math-ph/0010025}};
see also T.~Hahn,
    \textit{FeynArts}, \textit{FormCalc} and
    \textit{LoopTools} user's guides, available from \texttt{%
      http://www.feynarts.de}.



\bibitem{PDB2000}  D.E.~Groom \textit{et al.}, \textit{Review of Particle
      Physics}, \textit{Eur.\ Phys. J.\ }\textbf{C 15} (2000) 1.

 \bibitem{OPAL}  G. Abbiendi \textit{et al. }(OPAL Collab.), \textit{Eur.
      Phys. J.} \textbf{C 7} (1999) 407, \texttt{hep-ex/9811025};
 \textit{ibid.}, \textit{Eur. Phys. J.} \textbf{C 18} (2001) 425,
 \texttt{hep-ex/0007040}.

\bibitem{LEPHWG} ALEPH, DELPHI,L3, OPAL Collab. and the LEP Higgs
    Working Group, CERN-EP/2001-055, \texttt{hep-ex/0107030}.

\bibitem{Hollik}  M. Carena, H.E. Haber, S. Heinemeyer, W. Hollik, C.E.M.
    Wagner, G. Weiglein, \textit{Nucl. Phys. } \textbf{B 580} (2000)
    29, \texttt{hep-ph/0001002}.

\bibitem{Gambino} P. Gambino, M. Misiak, \textit{%
      Nucl. Phys. } \textbf{B 611} (2001) 338, \texttt{hep-ph/0104034}.

\bibitem{Spira} M. Spira, {HIGLU} and {HQQ} packages:
    \texttt{http://www.desy.de/$\sim$spira/higlu/}, and
    \\\texttt{http://www.desy.de/$\sim$spira/hqq/}.

 \bibitem{Spira2} M. Spira, \textit{Higgs and SUSY particle production
at hadron colliders} in: Proc. of SUSY 02 (DESY, Hamburg, 2002) p.
217, Eds. P. Nath, P.M. Zerwas, \texttt{hep-ph/0211145}.

\bibitem{Spira3} M. Spira, in: \textit{MSSM Higgs boson production at the LHC},
    proc. of the \textit{International Workshop on Quantum Effects in the MSSM},
    World Scientific 1998, p. 125, ed. J.
    Sol{{\`a}}, \texttt{hep-ph/9711407};
    M. Spira, \textit{Fortschr. Phys. } \textbf{46}
    (1998) 203, \texttt{hep-ph/9705337}.

\bibitem{CoarasaJS} J.A. Coarasa, R.A. Jim{\'e}nez, J. Sol{\`a},
\textit{Phys. Lett. } \textbf{B 389} (1996) 312, \texttt{hep-ph/9511402}.

\bibitem{ATLAS} ATLAS Collab., Detector and Physics Performance,
    Technical Design Report, CERN/LHC/99-14.

\bibitem{CMS} D. Denegri \textit{et al.}, Summary of the CMS
    Discovery Potential for the MSSM SUSY Higgses, CMS Note 2001/032,
    \texttt{hep-ph/0112045}.

\bibitem{Gianotti}  F.~Gianotti, \textit{Precision physics at the LHC},
    proc. of the \textit{IVth International Symposium on Radiative
      Corrections (RADCOR 98)}, World Scientific 1999, p. 270, ed. J.
    Sol{{\`a}}.

\bibitem{CarenaHaber}  H. Haber, \textit{Higgs theory and phenomenology in the standard model and MSSM}, in: Proc. of SUSY 02 (DESY, Hamburg,
2002) p. 58, Eds. P. Nath, P.M. Zerwas, \texttt{hep-ph/0212136};
M. Carena, H. Haber, \textit{Prog.\ Part.\ Nucl.\ Phys.}  {\bf 50} (2003) 63,
\texttt{hep-ph/0208209};
R. Kinnunen, \textit{Higgs physics at LHC}, J.F.\, Gunion,
\textit{Extended Higgs sectors}, \texttt{hep-ph/0212150},
in: Proc. of SUSY 02
(DESY, Hamburg, 2002) pp. 26,80, Eds. P. Nath, P.M. Zerwas.

\bibitem{Duncan} M.J. Duncan, \textit{Nucl. Phys. }\textbf{B 221}
    (1983) 285;
\textit{ibid.} \textit{Phys. Rev. }\textbf{D 31} (1985) 1139.

\bibitem{Besmer} T. Besmer, C. Greub, T. Hurth, \textit{%
      Nucl. Phys. } \textbf{B 609} (2001) 359, \texttt{hep-ph/0105292}.

\bibitem{Pokorsky} F. Gabbiani, E. Gabrielli, A. Masiero, L.
  Silverstrini, \textit{Nucl. Phys. }\textbf{B 477} (1996) 321,
  \texttt{hep-ph/9604387};
  M. Misiak, S. Pokorsky, J. Rosiek, in: Heavy Flavors II,
\textit{Adv.\ Ser.\ Direct.\ High Energy Phys.}  {\bf 15} (1998) 795,
ed. A.J. Buras, M. Lindner (World Scientific, Singapore, 1998),
\texttt{hep-ph/9703442}.

\bibitem{Siannah} J. Guasch, W. Hollik, S. Pe{\~n}aranda, \textit{Phys. Lett. }\textbf{B 515} (2001) 367, \texttt{hep-ph/0106027}.


\bibitem{bggs} A. Belyaev, D. Garcia, J. Guasch, J. Sol{\`a}, \textit{JHEP }
0206 (2002) 059, \texttt{hep-ph/0203031};
\textit{ibid.} \textit{Phys. Rev.
}\textbf{D 65} (2002) 031701, \texttt{hep-ph/0105053}.








\end{thebibliography}

%
%
%
%
%
%

\end{document}